\documentclass[pre,aps,superscriptaddress,amsmath,amssymb,twocolumn]{revtex4}
\usepackage{graphicx}
\renewcommand{\phi}{\varphi}
\begin{document}

\title{Equilibrium equation of state of a hard sphere binary mixture at very
large densities using replica exchange Monte-Carlo simulations}

\author{Gerardo Odriozola}  

\affiliation{Programa de Ingenier\'{\i}a Molecular, Instituto
Mexicano del Petr\'{o}leo, L\'{a}zaro C\'{a}rdenas 152, 07730
M\'{e}xico, D. F., M\'{e}xico}

\author{Ludovic Berthier}

\affiliation{Laboratoire des Collo{\"\i}des, Verres et
Nanomat{\'e}riaux, UMR CNRS 5587, Universit{\'e} Montpellier 2, 34095
Montpellier, France}

\date{\today}

\begin{abstract}
We use replica exchange Monte-Carlo simulations to measure the
equilibrium equation of state of the disordered fluid state for a binary hard
sphere mixture up to very large densities where standard Monte-Carlo
simulations do not easily reach thermal equilibrium. For the moderate
system sizes we use (up to $N=100$), we find no sign of a pressure
discontinuity near the location of dynamic glass singularities
extrapolated using either algebraic or simple exponential divergences, 
suggesting they do not correspond to genuine thermodynamic
glass transitions.
Several scenarios are proposed  for the fate of the fluid state in the
thermodynamic limit.
\end{abstract}


\maketitle

\section{Introduction \label{introduction}}

Simple liquids, crystals, glasses, powders, and colloidal dispersions are
frequently modeled using hard spheres. Although considered as one of the
simplest models in condensed matter physics, hard spheres exhibit a
complicated phase behavior that is not fully elucidated. In particular, a
well-established first order fluid-solid transition exists in three
dimensions. For monodisperse systems, it occurs from volume fractions
$\phi_{\rm f}= \pi \rho \sigma^3 /6$ $\approx$ $0.492$ to $\phi_{\rm
s} \approx 0.545$ ($\rho$ is the number density and $\sigma$ the particle
diameter) \cite{Alder57,Hoover68,Wilding00,Noya08,Odriozola09}. However, the
metastable fluid branch persists for $\phi > \phi_{\rm f}$, and its
fate at large $\varphi$ remains a debated subject \cite{Parisi10}. Since the
fluid cannot exist above the maximum density of the cubic centered crystal
structure, $\phi_{\rm fcc} \approx  0.74$, 
it may either go unstable, or it may exhibit a
singularity with a diverging (dimensionless) pressure, $Z = \beta P /\rho$,
and a vanishing (also dimensionless)
isothermal compressibility, $\chi=\delta\rho/\delta(\beta
P)$. Here $\beta=(k_B T)^{-1}$, with $T$ the temperature and $k_B$ the
Boltzmann constant. Additionally, a thermodynamic glass transition could
possibly occur along the 
way \cite{Parisi10,Speedy94,Yeo95,Blaaderen95,Speedy97,Foffi07},
characterized by a diverging timescale for structural 
relaxation~\cite{vanmegen1,vanmegen2,chaikin,gio}, a
change of slope in the equilibrium 
equation of state $Z(\phi)$, and a jump in the
compressibility. These features would be the analog, for hard spheres, of
the glass transition observed in glassforming liquids, characterized in
particular by a diverging viscosity and a jump in the specific 
heat~\cite{reviewnature}. It is
the aim of this work to search for a thermodynamic signature of the glass
transition in hard spheres.

Studying the metastable fluid branch by simulations is complicated 
since the system naturally tends to form the crystal
phase~\cite{Rintoul96a,Rintoul96b}, at least in
three dimensions \cite{charbonneau}.
Since pressure is very dependent on the existence of small amounts
of crystal nuclei, excluding ordered configurations from the sampling is
critical to obtain the real pressure-density relationship
\cite{Rintoul96a,Rintoul96b}. An efficient way to overcome the somewhat
arbitrary exclusion of crystalline states from the sampling is to introduce
size polydispersity to avoid, or at least considerably delay, crystal
formation. One must then 
work between several constraints: polydispersity must be
large enough to prevent ordering, but small enough that a 
qualitatively different physics, specific
to very polydisperse systems, does not set in. For instance, phase separation
can occur in mixtures~\cite{Biben91}, 
or fractionation in systems with continuous
polydispersity~\cite{Sollich10}. 
These phenomena have counterparts even for disordered states,
since multiple glass
transitions might occur in polydisperse systems, where for instance large
particles are arrested in a sea of small ones that still easily 
diffuse~\cite{donth95}. 
In this work we
use a 50:50 binary mixture of hard spheres with a diameter ratio 1.4,
large enough to efficiently prevent crystallization, but which shows no sign
of multiple glass transitions.

The final problem to be overcome is also the most difficult one: approaching
the glass transition at thermal equilibrium is hard  in systems where
the viscosity becomes large because the timescale to reach
equilibrium is simultaneously  
diverging. On this aspect, numerical simulations could 
potentially outperform experimental work since it is possible, at least in
principle, to imagine algorithms that have no `physical' counterpart but
still allow a proper exploration of the configuration space, and thus of the
thermodynamic properties of the system~\cite{Frenkel}. Several such `smart' 
algorithms exist in various context of statistical mechanics,
such as umbrella sampling which makes use of biased
statistical weights, replica exchange or parallel tempering 
where copies of the system
at various thermodynamic states are run in parallel to avoid being 
trapped in free energy minima \cite{replicarough,hukushima96,swendsen},
or cluster and swap algorithms which implement
unphysical particle moves to speed up 
equilibration~\cite{Krauth,Bernard09}. 

Although commonly used
and very successful in many areas of condensed matter, such methods have
comparatively been much less used in numerical studies of the glass
transition, for several reasons. Firstly, the glass transition 
is mostly defined by, and studied via, dynamic properties, 
and so it is vital to use physical microscopic dynamics, which
inevitably yields slow dynamics. 
There are nevertheless interesting thermodynamic
properties to be investigated in glassforming materials,
for which implementation of 
particle swaps \cite{grigera-parisi}, cluster moves \cite{Santen00},
Wang-Landau sampling \cite{wanglandau,depablo2}, 
or parallel tempering \cite{yamamoto-kob,sethna,sciortino,szamel}
have all been implemented. Of course 
these different methods can be combined to improve further
the efficiency. 
This has led in particular to strong claims 
about both the absence \cite{Santen00} and 
presence \cite{grigera-parisi,depablo} 
of thermodynamic glass transitions
in various glassy fluid models (including hard spheres), 
but also raised debates about the real efficiency
of the various numerical algorithms to study systems with slow 
dynamics \cite{sethna,yamamoto-kob,sciortino,dave}.

Here, we employ the replica exchange Monte-Carlo (REMC) method
\cite{Lyubartsev92,Marinari92,swendsen,hukushima96}, 
as recently adapted to systems 
composed of hard particles \cite{Odriozola09}. The idea is to 
simulate several replicas of the same system at
different but close enough thermodynamic states to allow efficient
exchanges between the replicas \cite{Frenkel}. For soft
interparticle potentials, the most common ensemble expansion is that
performed in temperature where each replica follows a canonical ensemble
simulation and the ensembles are set at different temperatures. 
To take advantage of the REMC algorithm for hard spheres, 
one needs to expand the isobaric-isothermal ensemble in 
pressure \cite{replicaisobaric},
each replica evolving at a different pressure \cite{Odriozola09}.
Replicas having the larger pressures can escape 
from locally stable free energy minima through successive exchanges with 
replicas at lower pressures \cite{Yan99}. 

Using REMC, we have been able to reach thermal equilibrium for hard spheres
up to very large densities where standard Monte-Carlo algorithms do not 
allow proper sampling of the configuration 
space~\cite{gio,szamel10,Berthier09}. 
We were thus 
able to study the thermodynamic properties of the disordered
fluid branch of a binary hard sphere mixture 
over a broad density range which includes both the mode-coupling,
$\phi_{\rm mct}$, and Vogel-Fulcher-Tamman, $\phi_{\rm vft} > 
\phi_{\rm mct}$, 
dynamic singularities finding no thermodynamic signature 
for any of them, at least for the moderate system sizes we used,
up to $N=100$.  
While the absence of a genuine transition at $\phi_{\rm mct}$
can be established by standard numerical 
methods~\cite{gio,Berthier09,szamel10}, 
it is the main
new result of this work that the same phenomenon seems to occur also 
at $\phi_{\rm vft}$.
 
The paper is organized as follows. 
In Sec. \ref{model} we describe in more detail 
the model we use, and review the various `critical' volume fractions
that have been reported in previous work.
In Sec.~\ref{method} we provide details about the REMC simulations.
In Sec.~\ref{test} we perform several tests to ensure that 
a proper sampling of configuration space has been done.
In Sec.~\ref{equilibrium} we describe our equilibrium results 
for the thermodynamics of the system.
In Sec.~\ref{nonequilibrium} we investigate even higher densities,
for which thermal equilibration could not be reached.
Finally, we discuss our results in Sec.~\ref{conclusion}.

\section{Critical densities in a binary hard sphere mixture model}
\label{model}

Previous work on binary mixtures suggests that a 50:50 binary
mixture of hard spheres with a diameter ratio of 1.4 is a very efficient way
to prevent crystalline ordering even at large densities 
\cite{Berthier09,harrowell,Speedy94}. We will use $N = N_A + N_B$ 
particles, $N_A$ and $N_B$ denoting the number of small and large 
particles in the mixture, respectively. We work in units where 
the diameter of the small particles is unity, $\sigma_{AA} = 1$. 

Moreover, the dynamics of small and large particles
is strongly coupled so that the slow relaxation and location of the 
putative dynamic glass singularities yields consistent results
for both components of the mixture~\cite{gio,gio2}. 
Thus, this model seems well-suited 
for investigating the existence of a thermodynamic glass transition 
of hard spheres. In a previous (bidimensional) 
study where efficient cluster Monte-Carlo moves were used~\cite{Santen00},
a very large polydispersity was introduced, with the 
unwanted result that large particles seemed to arrest at a density where
small particles could still easily diffuse, making the identification
of dynamic singularities somewhat ambiguous~\cite{dave}.

Previous numerical explorations of the dynamics of the present 
binary mixture revealed the existence of very slow dynamics and possible
dynamic singularities at large volume fraction, with no interference
from the crystalline phase~\cite{gio}. Several relevant values of the 
packing fractions have been reported using different definitions
and theoretical approaches, 
and we summarize them in Table \ref{table}.

\begin{table}
\begin{tabular}{| l | l |}
\hline
Definition  &  Volume fraction     \\
\hline 
Onset of glassy dynamics & $\phi_{\rm onset} \approx 0.56$ \\
Mode-coupling theory, Eq.~(\ref{mct})  &  $\phi_{\rm mct} = 0.592$    \\
Vogel-Fulcher-Tamman, Eq.~(\ref{vft})  &  $\phi_{\rm vft} = 0.615$    \\
Dynamic scaling, Eq.~(\ref{scaling})       &  $\phi_0 = 0.635$    \\
Diverging pressure (lower bound)  &  $\phi_{\rm low} = 0.662$    \\  
\hline
\end{tabular}
\caption{\label{table} Values of the relevant volume fractions 
characterizing the physical behaviour of the fluid for
the binary hard sphere mixture studied in this work.}
\end{table}

First, the dynamics of the system slows down and starts to become
non-exponential above $\phi_{\rm onset} \approx 0.56$, which can thus be seen
as the onset density for slow dynamics in this system.

Second, the location of several
dynamic `singularities' can be defined and have been numerically studied. 
An algebraic divergence of the relaxation time, 
\begin{equation}
\tau \sim (\varphi_{\rm mct} - \varphi)^{-\gamma},
\label{mct}
\end{equation} 
as predicted by mode-coupling theory~\cite{gotzebook}, 
can be located near $\varphi_{\rm mct}
\approx 0.592$.  However, simulations also revealed this density to be 
a crossover since the equilibrium relaxation time can be measured 
at and above  $\phi_{\rm mct}$ where it remains 
finite~\cite{gio,gio2}. 
This suggests that a different functional form should 
be used to extrapolate a possible divergence of the 
relaxation time.

A popular functional form for $\tau(\phi)$ is the 
so-called Vogel-Fulcher-Tamman (VFT) expression~\cite{reviewnature}, 
\begin{equation}
\tau \sim \tau_\infty
\exp \left( \frac{A}{\varphi_{\rm vft} - \varphi} \right),
\label{vft}
\end{equation} 
which yields,
for the present system, the value $\varphi_{\rm vft} \approx
0.615$, $A$ and $\tau_\infty$ being additional fitting parameters~\cite{gio}.
As opposed to the mode-coupling singularity, standard simulations
fail to access such a large packing fraction in equilibrium conditions, 
since the largest 
state point investigated in Ref.~\cite{gio} 
is $\phi = 0.597 < \phi_{\rm vft}$.  
  
Using a combination of scaling arguments involving both 
direct simulations of hard particles, and a soft harmonic 
repulsion at very low temperatures, recent numerical work 
provided support for the existence
of a slightly different, stronger dynamic 
divergence~\cite{gio,gio2,Berthier09,Berthier09b}, 
\begin{equation}
\tau \sim  \tau_\infty \left( \frac{A}{(\varphi_{0} - \varphi)^\delta} 
\right)
\label{scaling}
\end{equation}
with the preferred values $\delta \approx 2.2$ and $\phi_0
\approx 0.635$.

Finally, taking the view that no thermodynamic glass transition occurs, 
one must conclude that dynamics should arrest when particles come
into contact and no particle move can take place. In this perspective, 
$\tau$ must diverge simultaneously with the pressure $Z$ at the 
random close packing or jamming density, $\phi_{\rm rcp}$, 
which can then be  
empirically defined as the end point of the equilibrium 
equation of state of the fluid branch \cite{Kamien07}. 
In practice this is hard to measure
because the system falls out of equilibrium and becomes a nonergodic 
hard sphere glass much before getting to jamming,  
such that only lower bounds to the location of the diverging pressure
can be numerically determined. For the present system, 
previous work reported the value $\phi_{\rm low} \approx 0.662$ as the tightest
lower bound on $\phi_{\rm rcp}$, obtained by rapid compressions
of carefully equilibrated fluid states~\cite{Berthier09,pinaki}. 
This result indicates that
the putative end point of the metastable fluid branch for this system
is above $\phi_{\rm low} = 0.662$.

\section{The replica exchange Monte-Carlo method}
\label{method}

The partition function in the extended ensemble studied 
in the replica exchange Monte-Carlo method we use 
is given by \cite{replicaisobaric,Odriozola09}
\begin{equation}
Q_{\rm extended}=\prod_{i=1}^{n_r} Q_{N T P_i},
\end{equation} 
where $Q_{NTP_i}$ is the partition
function of the isobaric-isothermal ensemble of the system at pressure
$P_i$, temperature $T$, particle number $N$. The important new  
parameter is $n_r$, the considered number of replicas of the system. 

This extended ensemble is sampled by
combining standard $NTP_i$ simulations on each replica (involving both
trial displacements of single particles and trial volume changes) 
and replica exchanges (swap moves at the replica level). 
To satisfy detailed balance, these swap moves are 
performed by setting equal all a
priory probabilities for choosing adjacent pairs of replicas
and using the following acceptance 
probability~\cite{replicaisobaric,Odriozola09}
\begin{equation}
\label{accP} 
P_{\rm acc}\!=\! \min(1,\exp[\beta(P_i-
P_j)(V_i-V_j)]), 
\end{equation} 
where $V_i-V_j$ is the volume difference
between replicas $i$ and $j$. Adjacent pressures should be close enough to
provide nonnegligible 
exchange acceptance rates between neighboring ensembles. In
order to take good advantage of the method, the ensemble at the smaller
pressure must also 
ensure large jumps in configuration space, so that the larger
pressure ensembles can be efficiently sampled.

The probability for selecting a particle displacement trial, $P_d$, for
selecting a volume change trial, $P_v$, and a swap trial, $P_s$, are fixed
to 
\begin{equation}
\label{PdPvPs} 
\begin{array}{lll} 
P_d & = &
n_rN/(n_r(N+1)+w), \\ P_v & = & n_r/(n_r(N+1)+w), 
\\ P_s & = & w/(n_r(N+1)+w),
\\ 
\end{array} 
\end{equation} 
where $w \ll 1$ is a weight factor. Note that
$P_d+P_v+P_s=1$, as it should. The
probability density function to have the next swap trial move at the trial
$n_t$ is given by 
\begin{equation}
\label{Pdfs} P(n_t) = P_s \exp(-P_s n_t). 
\end{equation} Hence, one may obtain the next swap trial move from
$n_t= - \ln(\xi) / P_s$, with 
$\xi$ being a random number uniformly distributed
in the interval $]0,1]$ \cite{Gillespie77,Odriozola03}.  
We set all particles of a given replica to have the same a
priori probability of being selected to perform a displacement trial.  The
same is true for selecting a replica for performing a volume change trial.

The trials $[1,n_t-1]$ are
displacements and volume changes, and so, they can be independently
performed on the replicas. This has the advantage of being easily
parallelized. 
The algorithm is parallelized in
four threads, since quad core desktops are used, but could be 
more efficiently parallelized in
$n_r$ threads.  
Since all swap trials are performed in a single core, the efficiency
of the parallelization increases with decreasing $w$. We employed $w=1/100$.
Verlet lists are used for saving CPU time, which can be 
quite large for the replicas evolving with the highest pressure values.
 
Our simulations are performed in two steps. 
All simulations are started by randomly placing particles
(avoiding overlaps), so that the initial volume fraction is $\varphi=0.30$.
We first perform about $2 \times 10^{13}$ 
trial moves at the desired state points, during 
which we observe that the replicas reach a stationary state. 
We then perform  more $2 \times 10^{13}$  additional trials during which
various measurements are performed, with results described in the following
sections.

The maximum particle displacements and volume changes for trial moves 
are adapted for each pressure to yield acceptance rates close to 0.3. 
Thus, particle displacements and volume changes of ensembles 
having high pressures are smaller than those associated to ensembles 
having low pressures. 
An optimal allocation of the replicas should lead to a constant
swap acceptance rate for all pairs of adjacent ensembles. For a temperature
expansion, the efficiency of the method peaks at swap acceptance rates close
to 20\% \cite{Rathore05}.  In this work, we use 
instead a geometric progression
of the pressure with the replica index. 
In Sec.~\ref{equilibrium} we report results for various 
system sizes, $N=60$, 80, and 100 using $n_r=14$, with 
$\beta P$ varying from 38 to approximately 5.8, the geometrical 
factor being 0.865.
In Sec.~\ref{nonequilibrium} we present additional results 
where the largest pressure 
is $\beta P = 100$, $N=60$, $n_r=18$, and the geometrical factor 
is 0.840. 

\section{Thermalization tests}
\label{test}

\begin{figure} 
\resizebox{0.45\textwidth}{!}{\includegraphics{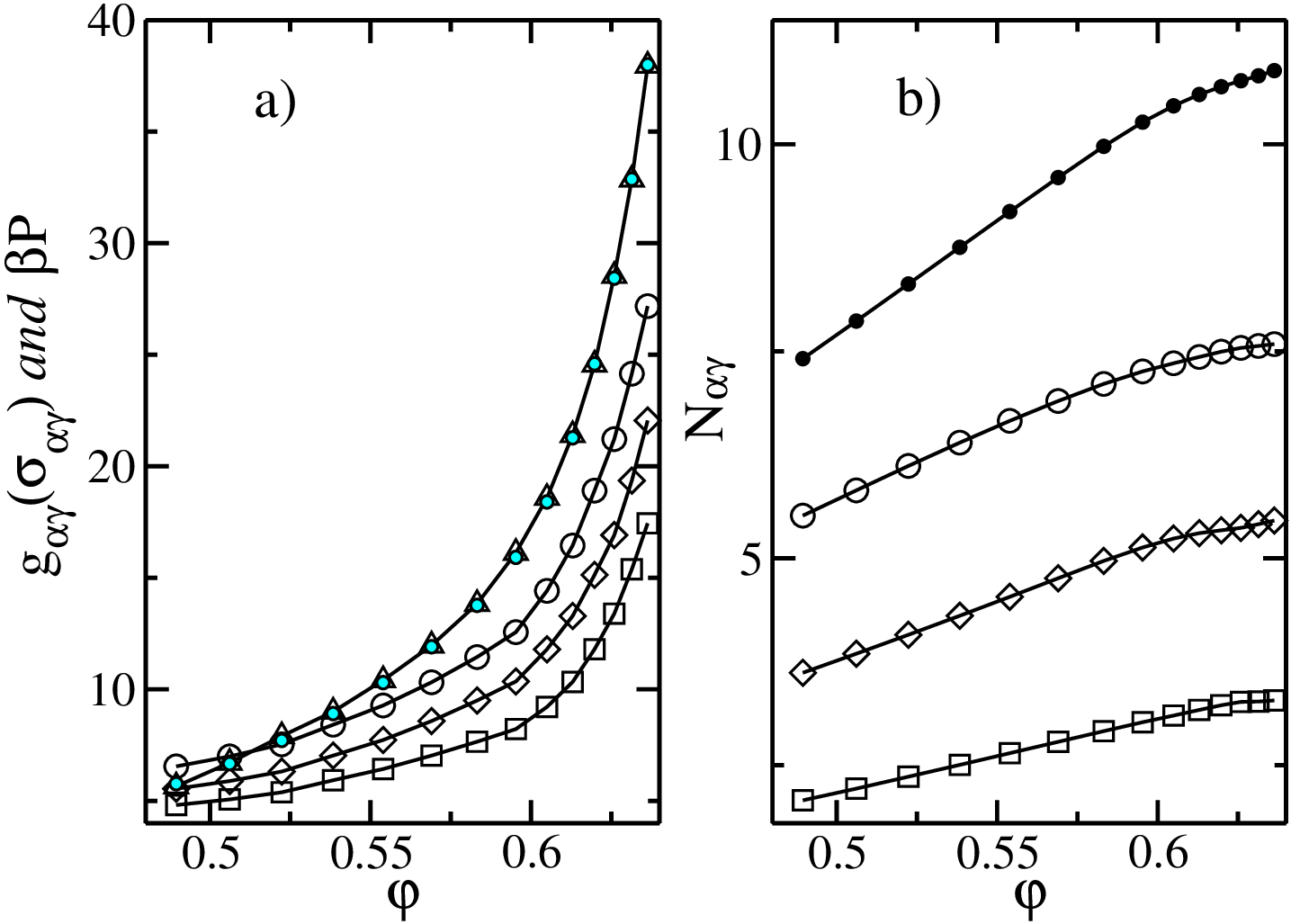}}
\caption{\label{Press-bond} 
a) $g_{AA}(\sigma_{AA})$ (squares), $g_{AB}
(\sigma_{AB})$ (diamonds), and $g_{BB}(\sigma_{BB})$ (circles) as a function
of $\varphi$. The total pressure, Eq.~(\ref{eqP}), is
shown as triangles. It agrees well with the set pressure 
values (light bullets). b) The
number of AA (squares), AB (diamonds), and BB (circles) neighbors
as a function of $\varphi$ with bullets indicating the total
number of neighbors per particle. All data correspond to $N=100$.} 
\end{figure}

The aim of this work is to provide new, reliable thermodynamic 
information at large densities where thermalization becomes a severe 
issue for standard algorithms.
This means in particular that the algorithm must be able to 
sample accurately a phase space where ergodicity is potentially
broken in the thermodynamic limit. Such severe sampling conditions
are also met for instance in systems such as spin glasses \cite{Marinari92}.
It is crucial to establish whether the produced results 
are indeed representative of thermal equilibrium, as we now discuss. 

As a first check we verify that the pressure
measured from the configurations sampled by the replicas
in each  ensemble yield results consistent with the
values set numerically. Pressure and structure are related by~\cite{Biben91}
\begin{equation}
\label{eqP} 
\frac{\beta P}{\rho}=1+\frac{2 \pi \rho}{3}
\sum_{\alpha} \sum_{\gamma} x_{\alpha} x_{\gamma} \sigma^3_{\alpha \gamma}
g_{\alpha \gamma}( \sigma_{\alpha \gamma}) 
\end{equation} 
where $\alpha$
and $\gamma$ run over species $A$ and $B$, and $x_{\alpha}$,
$\sigma_{\alpha \gamma}$, and $g_{\alpha \gamma}$ respectively 
being the fraction of particles
in species $\alpha$, the contact distance between $\alpha$ and
$\gamma$, and the partial radial distribution functions of species $\alpha$ and
$\gamma$.  Note that $g_{\alpha \gamma}(\sigma_{\alpha \gamma})$ 
must be evaluated using
a careful extrapolation of $g_{\alpha \gamma}(r)$ towards contact. Thus, we
may split the excess pressure into three contributions, corresponding to the
AA, AB, and BB interactions.
These contributions are shown in
Fig.~\ref{Press-bond}-a together with
the total pressure obtained from Eq.~(\ref{eqP}).
As can be seen, the measured pressure agrees very well 
with the values imposed numerically. Furthermore, a
smooth behavior is obtained for all $g_{\alpha \gamma} (\sigma_{\alpha
\gamma})$ as a function of $\varphi$ suggesting that 
adequate sampling has been performed. 

\begin{figure} 
\resizebox{0.45\textwidth}{!}{\includegraphics{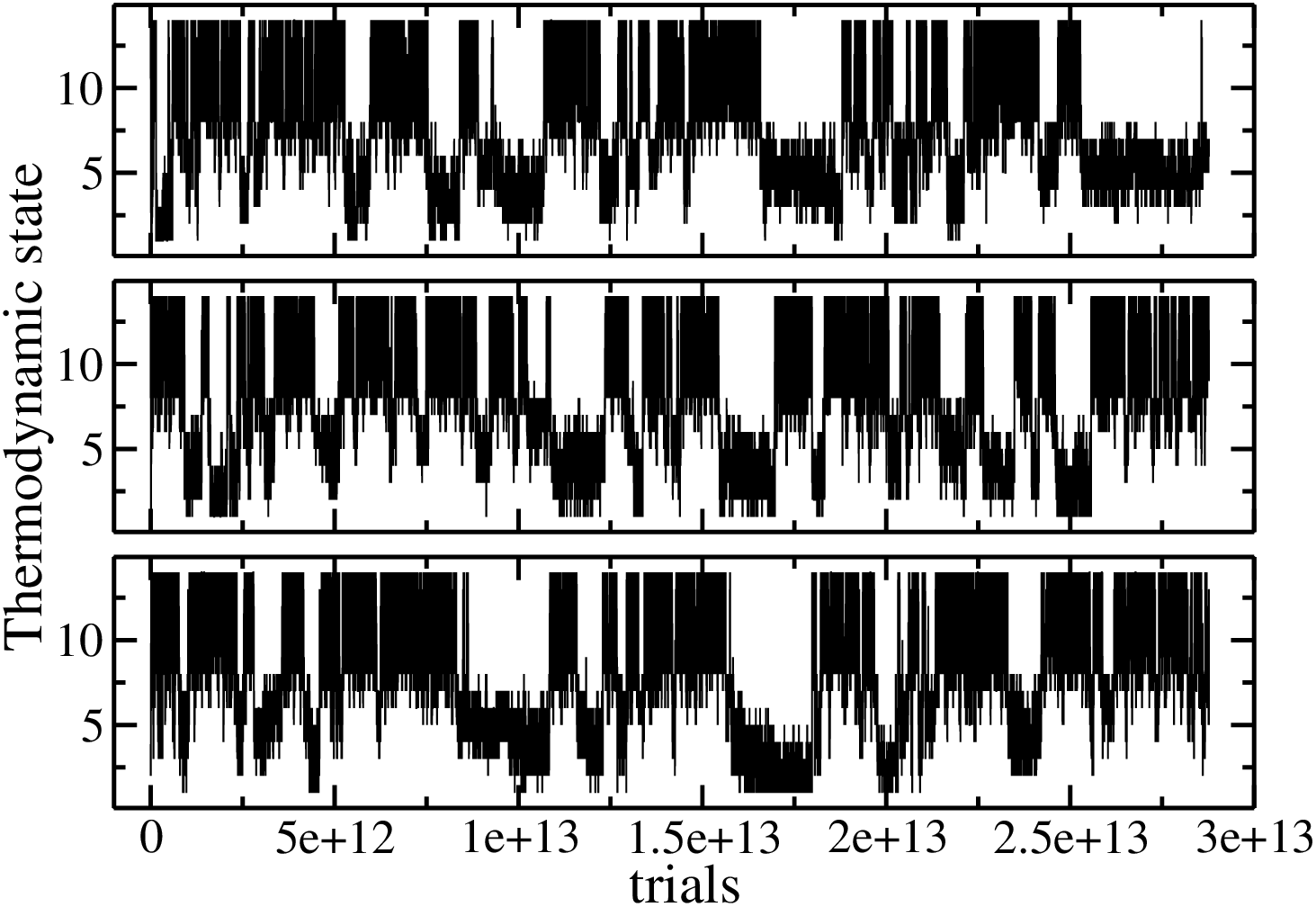}}
\caption{\label{Swap} 
Random walk in pressure space for $N=100$ and $n_r=14$
for three arbitrarily chosen replicas.
Thermodynamic state `1' corresponds to the highest
pressure and `14' to the lowest. All replica visit several
times both lowest and highest pressure states.}
\end{figure}

For all $\varphi$ the largest
contribution to the excess pressure is that of the large-large pairs
(BB), followed by the large-small (AB) and the small-small (AA)
pairs, in that order. In the right panel of Fig.~\ref{Press-bond} 
the evolution of the average number of neighbors
is shown, obtained by integration of the partial pair correlation
functions in a spherical shell of constant 
thickness $0.2\sigma_{AA}$. The total 
number of neighbors per particle is also shown.
The numbers of neighbors are consistent with the contributions to
the excess pressure, i.~e., they increase following the order AA, AB,
and BB at all $\varphi$ and they increasing with
$\varphi$, albeit more slowly than the pressure. This saturation 
is physically expected 
since the number of neighbors remains finite even 
when the pressure diverges near jamming.

We mentioned in the introduction that preventing crystallization
is in principle dealt with by using a binary mixture. Since evidence
for this stems from standard numerical approaches, it remains to be seen
whether REMC finds more easily the crystalline phase or not. 
Indeed previous work on the monodisperse system
showed that the REMC algorithm is not only capable
of forming the crystal phase but also to accurately predict the liquid-solid
transition \cite{Odriozola09}. Thus, if a crystal is the preferred state, we
expect to see signs of local orientational order. 
We checked this by computing the well-known order parameter 
$Q_6$, as defined for instance in 
Refs.~\cite{Steinhardt83,Rintoul96b,Odriozola09}, which is 
very sensitive to any trace of local angular order \cite{Rintoul96b}. We
evaluate it separately for AA, BB, and AB pairs. In all cases  and
at all densities, $Q_6$ is very close to the value of a
completely random system of points \cite{Rintoul96b}. Moreover, the three
$Q_6$ values do not evolve significantly during the runs. 
Thus, we can safely conclude that if the crystal 
phase corresponds to the equilibrium state of  this particular 
binary mixture, it is sufficiently metastable not to affect our results
regarding the disordered state.

\begin{figure} 
\resizebox{0.45\textwidth}{!}{\includegraphics{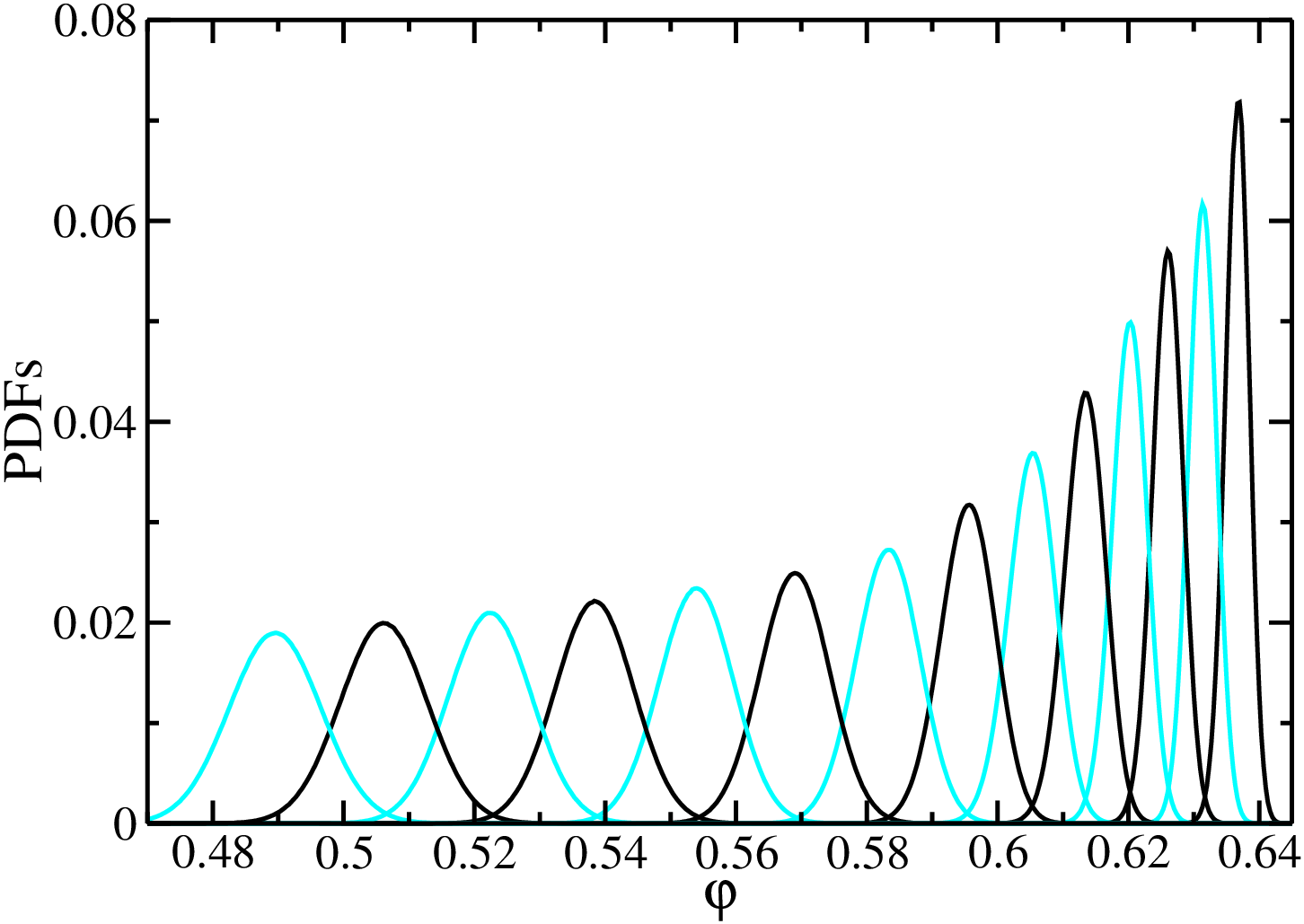}}
\caption{\label{PDFs} 
Probability distribution functions (PDFs) of volume fraction 
fluctuations for each of the $n_r=14$ pressure values, $N=100$.
All distributions are stationary, featureless, and symmetric
and have sufficient overlap to allow for replica exchanges.} 
\end{figure}

For replica exchange methods to provide an efficient sampling
of phase space, it is important to check whether all simulated replicas
visit the entire set of thermodynamic conditions several times. 
This is a necessary condition for thermalization because this ensures
that the configurations contributing to the thermodynamic
averages are very different as the low pressure replicas 
evolve rapidly and have large displacements in configuration space.   
In Fig.~\ref{Swap} we show the evolution of three randomly selected 
replicas making a random walk among the different pressure states.
We observe that all replicas contribute several times during the course
of the production run to both the highest and the lowest pressure 
states. With the imposed geometric progression of the pressure, we find 
that the acceptance rate has a small drop near $\phi \approx 0.58$,
which suggests that thermalization becomes much harder above these 
densities. It is intriguing that this corresponds roughly 
to $\phi_{\rm mct}$, above which it also becomes hard to reach thermal 
equilibrium using standard numerical tools. 
This numerical bottleneck suggests a faster decrease
of the number of accessible configurations as pressure increases, 
or a faster increase of the barriers separating long-lived 
metastable states.

Each replica evolves at a given predefined pressure $P_i$. 
Therefore, the volume $V_i$ of replica $i$ is a fluctuating quantity,  
and it is interesting to focus on the 
probability density functions (PDFs) of the volumes $V_i$, 
or equivalently of the corresponding volume fractions $\phi_i$. 
In cases where the system remains trapped in long-lived metastable states, 
the PDFs may be distorted or may contain peaks or shoulders which 
help detecting a lack of thermalization.
Additionally, these features of the PDFs typically disappear as time 
increases and thus help revealing whether measurements are performed in 
stationary states.
We observe that the PDFs evolve in the first simulation steps
but they then become both symmetric and take a Gaussian
shape. The resulting functions are shown in Fig.~\ref{PDFs} for
$N=100$ with low volume fractions PDFs correspond to low pressures. 
Notice that the PDFs have a larger peak and become narrower
as pressure increases, which reflects the fact that 
the compressibility decreases.  

\begin{figure} 
\resizebox{0.45\textwidth}{!}{\includegraphics{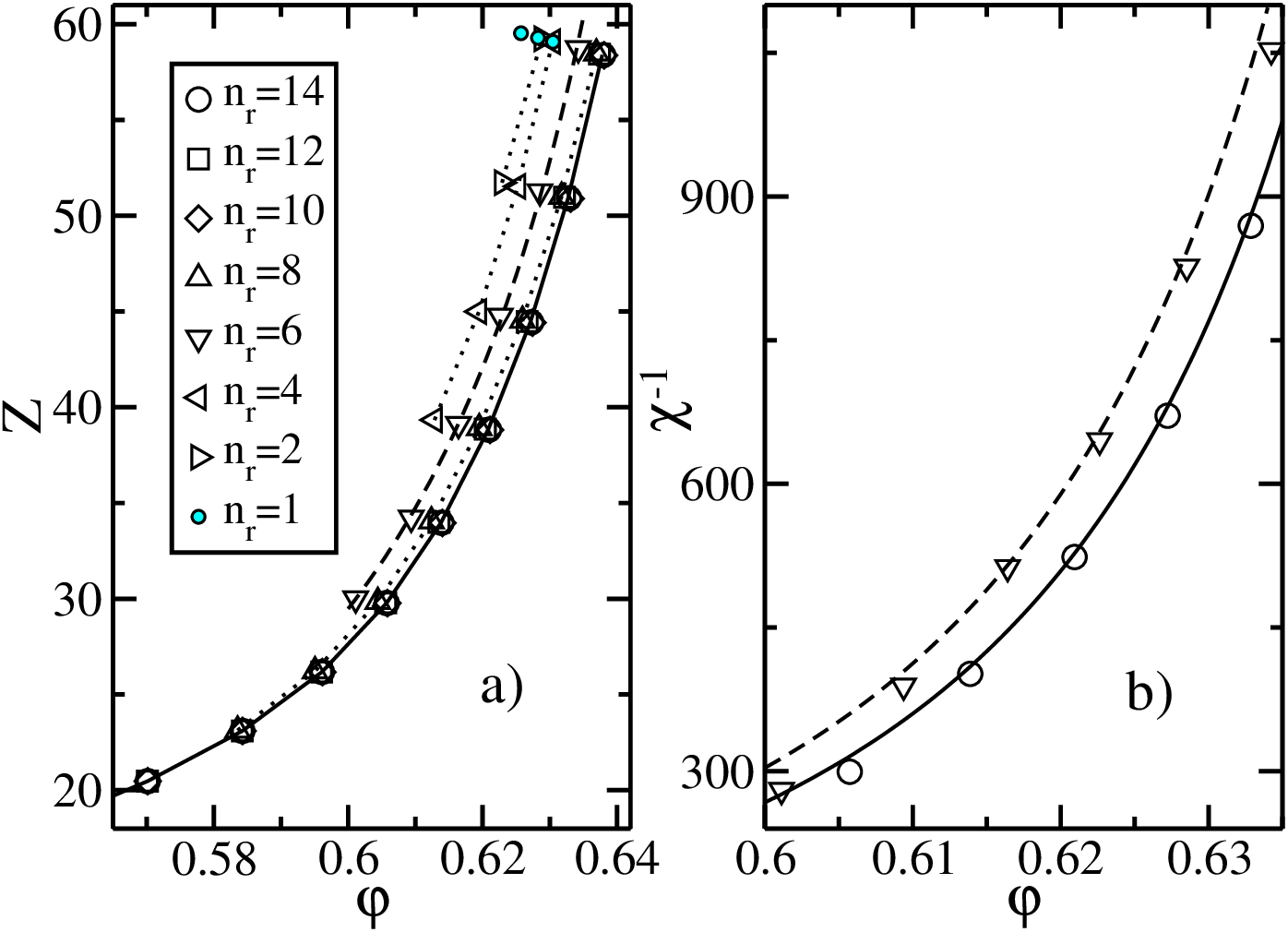}}
\caption{\label{Rep} 
a) Equation of state, $Z=\beta P/\rho = Z(\phi)$,
for $N=60$ and increasing number of replicas 
from $n_r=1$ to $n_r=14$. Only data for $n_r>8$
are reproducible, while data for a smaller $n_r$ are not thermalized.
b) Checking the FDT relation, Eq.~(\ref{FDT}), for density
fluctuations. The lines show $1/\chi$ obtained from 
taking the derivative of the equation of state, while symbols 
are direct evaluation using spontaneous fluctuations of the density.
The FDT holds with good accuracy both for equilibrated
systems ($n_r=14$, bottom) and for nearly frozen ones ($n_r=6$, top).}   
\end{figure}

Ergodicity implies that the same results should be obtained 
independently of the set initial conditions and of the 
parameters of the simulation.
We checked the reproducibility of our results by running 
simulations with $n_r=$ 1, 2, 4, 6, 8, 10, 12, and 14, for 
$N=60$, for the same highest pressure ($\beta
P=38$) and geometrical factor of 0.865. Three independent runs
were carried out with $n_r=1$, having all different initial conditions. 
For a fair comparison, 
all simulations lasted four weeks
running on identical single cores and no parallelization 
was implemented for this particular 
test. From the measured PDFs at each pressure, 
we measure the averaged density to calculate the equation of state
$Z(\phi)$, which are reported in Fig.~\ref{Rep}. We observe that results
become reproducible only when $n_r \geq 8$, which corresponds 
to simulations where the lowest pressure is below 
$\varphi = 0.58$ and yields thermalized results. For smaller $n_r$, $Z$ is
always larger than that obtained for $n_r \geq 8$, suggesting 
that thermal equilibrium had not been reached.
In particular, the three
independent runs with $n_r=1$ are well above the equilibrated curve and 
distinct from one another.
This implies that runs with $n_r < 8$ are nonergodic and do not sample
the configuration space accurately at large densities even with a large number
of trials. In particular, this means that a standard Monte-Carlo 
algorithm would not yield equilibrium results at large density, 
and that it is clearly outperformed by the REMC simulation scheme
we use in this work. 

A final test for equilibrium was suggested by Santen and Krauth
\cite{Santen00}. At thermal equilibrium the spontaneous 
fluctuations of density are related to the isothermal compressibility,
which is defined from the response of the pressure to an 
infinitesimal change
in density in the linear response regime. 
This relation is thus a form of  fluctuation-dissipation 
theorem (FDT), 
which is derived using the hypothesis that
states are sampled with the equilibrium Gibbs measure:
\begin{equation}
\chi = N \left( 
\frac{\langle\rho^2\rangle-\langle\rho\rangle^2}{ \langle\rho\rangle^2 }  
\right) = \frac{\delta\rho}{\delta(\beta P)}.
\label{FDT}
\end{equation}
Checking whether this relation is satisfied by the data 
is therefore in principle a good way to check thermalisation.
In practice, this means checking the existence of a quantitative
relationship between the broadness of the PDFs in Fig.~\ref{PDFs}
and the location of their averages. 

We followed these two routes for obtaining $\chi$. The fluctuations 
are directly measured from the simulations, while the response function
is obtained by first fitting the pressure locally to a smooth polynomial
function before taking the derivative with respect to density.
In Fig.~\ref{Rep} we present our results showing $1/\chi$ as a function
of density using both methods. When $n_r=14$ and results are reproducible,
we find that the FDT relative to fluctuation density is well satisfied, 
which comes as an additional proof that our data are 
representative of thermal equilibrium. 
However we note that for runs with a small number of replica 
all concentrated in the high density regime which appeared 
far from equilibrium in the left panel of Fig.~\ref{Rep}, 
the FDT is also satisfied with a good accuracy.
In that case, all replicas belong to the glassy state and are
basically frozen in a single `basin' where they sample 
quasi-equilibrium short-lived fluctuations. 
Studies of FDT violations in aging glasses have indeed shown
that deviations from FDT appear only when considering those
degrees of freedom that relax very slowly in the glass~\cite{CKP97}.   
This suggests that the FDT test suggested in Ref.~\cite{Santen00} 
is only effective in a narrow density regime where 
a complete separation of timescale does not make 
Eq.~(\ref{FDT}) valid even very far from equilibrium.
That is, the test seems useful to detect a slow evolution and thus, the
consistency of both $\chi$ determinations 
only guarantees a stationary state has been reached. 
This is a necessary but insufficient condition for equilibrium.

\section{Thermodynamic results at equilibrium}
\label{equilibrium}

In previous sections, we provided evidence that the REMC algorithm
is properly implemented, and that it might give thermalized results 
for $N=100$, $n_r=14$ up the pressure $\beta P = 38$. In this section
we study more carefully the outcome of this study, starting with 
the equation of state $Z(\phi)$.

\begin{figure} 
\resizebox{0.48\textwidth}{!}{\includegraphics{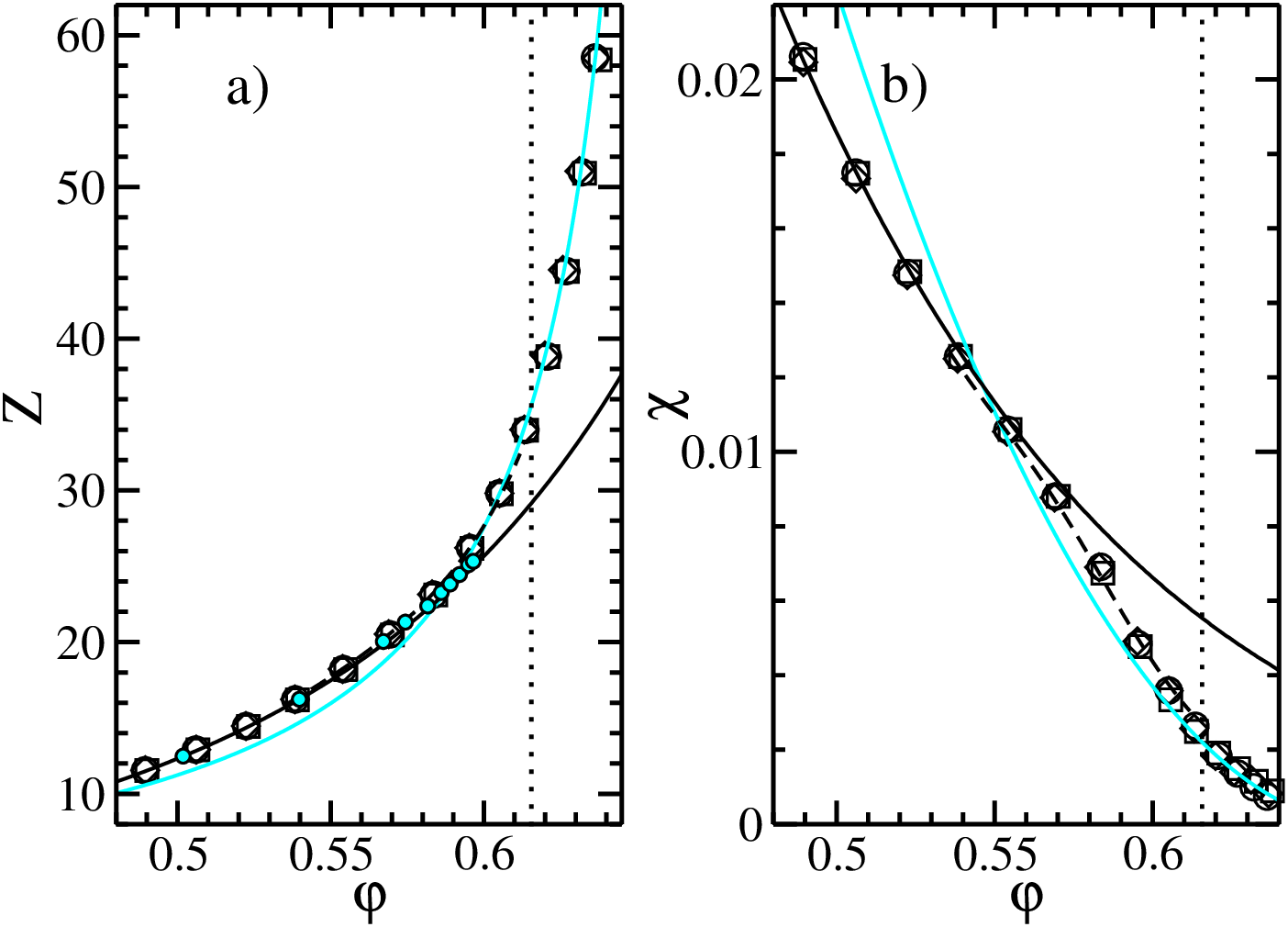}}
\caption{\label{Zrho-chi} 
a- Equation of state $Z(\phi)$ for different system sizes, 
$N=60$ (open squares), 
$N=80$ (circles), and $N=100$ (diamonds).
Light bullets are $N=1000$ data taken from Ref.~\cite{Berthier09}.
The black solid line is the BMCSL equation of state,
the dashed line an empirical polynomial form, and 
the light line a simple pole divergence, Eq.~(\ref{freefit}).
b- Isothermal compressibility obtained from density fluctuations
(symbols) or by derivative of the fits shown in panel a. 
The vertical dotted line is at $\phi_{\rm vft}$, where no thermodynamic
signature of a glass transition is found.}
\end{figure}

Using the PDFs shown in Fig.~\ref{PDFs} it is easy to deduce 
the average volume fraction for each pressure, and thus to obtain 
$Z(\phi)$. The results are shown in Fig.~\ref{Zrho-chi}-a for three 
different systems sizes $N=60$, 80 and 100. A comparison of the 
three system sizes shows that finite size effects appear to be 
very small for the equation of state as the data obtained with
different system sizes practically coincide. Nevertheless, larger system
sizes produce a very small but apparently 
systematic decrease of the volume fraction
at a given pressure, for all $\varphi$. As a further check, we report 
the results of an independent study using a standard Monte-Carlo
approach which used $N=1000$, but covers a smaller range of 
pressures~\cite{Berthier09}. Up to $\phi \approx 0.595$ where both data
sets can be compared, the data agree very well, confirming the
validity of our algorithm, at least up to this density.

A first quantitative 
result from our study stems from data at volume fractions larger than
the ones studied in Ref.~\cite{Berthier09}, which were all accurately
described using the Boublik-Mansoori-Carnahan-Starling-Leland (BMCSL) 
equation of state~\cite{Boublik70,Mansoori71}, 
which is the extension for mixtures
of the Carnahan-Starling equation of state. The data in 
Fig.~\ref{Zrho-chi}-a follow the BMCSL equation up to $\phi\approx 0.59$,
but clearly deviate from it at larger volume fractions, the deviations
becoming very large at large $\phi$ where BMCSL clearly underestimates
the pressure. A similar deviation was recently reported
from molecular dynamics simulations of a hard sphere system
with continuous polydispersity~\cite{hermes}.

As a better description of the data at large $\phi$ we used 
the fitting formula suggested from free volume 
considerations~\cite{freevol}, 
\begin{equation}
Z = \frac{d' \phi_c}{\phi_c - \phi},
\label{freefit}
\end{equation}
where $d'$ and $\phi_c$ are free fitting parameters. 
Although the prefactor $d'$ should in principle be constrained 
within free volume theory to be equal to the spatial dimension $d$, 
we find that its value must 
be adjusted to describe our data. In Fig.~\ref{Zrho-chi}-a 
we show the best fit to the data $\phi > 0.61$ to Eq.~(\ref{freefit})
as a full line, using $d'=2.82$ and $\phi_c = 0.669$. 
This value of $\varphi_c$ should be compared to the 
lower bound for the diverging pressure discussed 
above in Sec.~\ref{model} (see Table \ref{table}) which 
was $\phi_{\rm low}=0.662$~\cite{pinaki}. 
The large difference between the two values
is a direct sign that the REMC algorithm has been able to 
thermalize the system much more efficiently.
Note also that the fit in Eq.~(\ref{freefit}) only works
at large volume fractions, while at low $\phi$
values it clearly deviates from the simulation data.

Finally to account for the crossover region
$\phi \approx 0.58 - 0.61$ between the BMCSL and free volume
fits, we use an empirical high order polynomial fit, shown with a
dashed line. We give no particular emphasis on a physical interpretation
of this fit, which we simply use as a fitting tool to
obtain the numerical derivative of the pressure, and thus
the compressibility, in this intermediate
Range of volume fractions.    

A second important result of our study is obtained by considering the
vertical line which corresponds to the volume fraction
$\varphi_{\rm vft}=0.615$. While an extrapolation of the
relaxation time divergence using Eq.~(\ref{vft}) indicates the 
possibility of a glass transition occurring at 
$\varphi_{\rm vft}$, there is no corresponding thermodynamic 
signature in the equation of state, in particular no sign that 
a kink develops as the system size increases, at least for the 
modest $N$ values we have been able to study, see Sec.~\ref{conclusion}.

These results are confirmed in  Fig.~\ref{Zrho-chi}-b
which shows the evolution of the isothermal compressibility 
as a function of $\phi$ for the different system sizes.
These data are directly obtained from the spontaneous 
density fluctuations, i.~e., from
$\chi=N(\langle\rho^2\rangle-\langle\rho\rangle^2)/\langle\rho\rangle^2$, 
and they directly confirm the absence of any jump in the compressibility
over this range of volume fractions and system sizes, 
in particular near $\phi_{\rm vft}$. 

We also show in this figure the compressibility values 
as obtained from
$\delta\rho/\delta(Z\rho)$, using the three fits described above, namely
using the BMCSL equation of state at low $\phi$ (black
line), the polynomial fit at intermediate $\phi$ (dashed line)
and the free volume fit at high $\phi$ (light line). There
is excellent agreement between both sets of data showing that
Eq.~(\ref{FDT}) is well satified at over the entire range of
volume fractions. 

The compressibility data simply amplify the 
results obtained for $Z(\phi)$. In particular, 
the good agreement between the BMCSL equation of state 
and the numerical data is very good up to $\varphi \approx 0.56$, 
but deviations in fact 
already appear at moderate volume fractions $\phi \approx 
0.57-0.59$, that are not obvious from the pressure itself
(see Fig.~\ref{Zrho-chi}-a).
Similarly, the free volume description of the data
is only adequate above $\phi=0.61$.
These two limits make evident 
the existence of a crossover regime $\phi \sim 0.56-0.61$
where neither approaches work, and the only description 
we have is an empirical polynomial function, 
which, interestingly, shows two changes of the concavity
but no jump.

To sum up, our simulation data at low and
intermediate densities agree with the BMCSL equation of state, while at
large densities they are much better described by a simple divergence
at $\varphi_c=0.669$. The data show no jump of
$\chi$ in the studied range of $\varphi$ values, which encompasses
both $\phi_{\rm mct}$ and $\phi_{\rm vft}$
fitted dynamic singularities.

\section{Increasing the pressure further: nonequilibrium effects}
\label{nonequilibrium}

\begin{figure} 
\resizebox{0.48\textwidth}{!}{\includegraphics{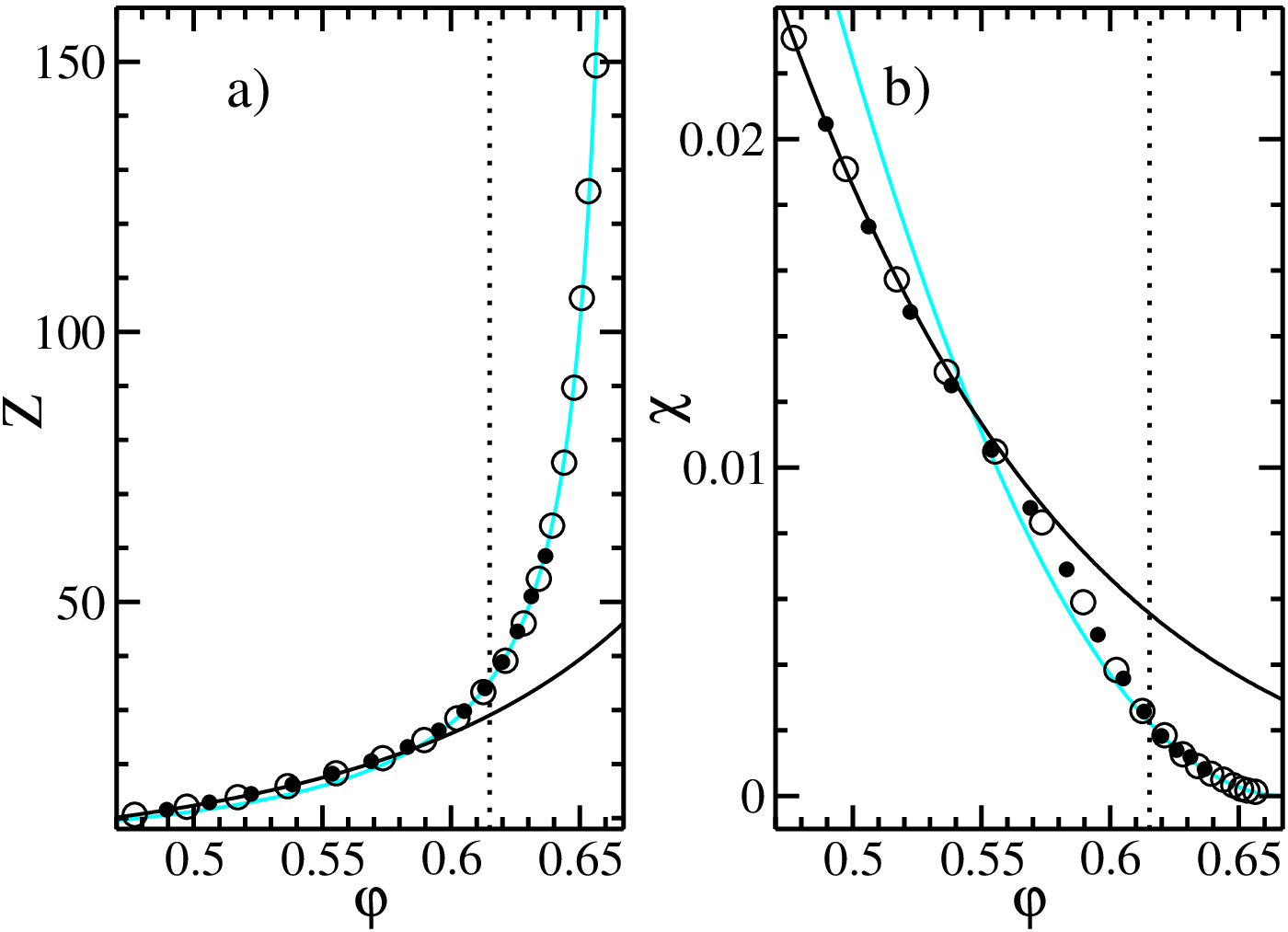}}
\caption{ \label{Z-chi-long} 
Same as Fig.~\ref{Zrho-chi} comparing runs with ($N=100$, $n_r=14$, 
maximum pressure $\beta P = 38$, filled circles)
to ($N=60$, $n_r=18$, maximum pressure $\beta P = 100$, open circles).  
Black lines correspond
to the BMCSL equation of state, light lines to free volume fit
Eq.~(\ref{freefit}), the vertical line denotes 
$\varphi_{\rm vft}=0.615$. While both data sets coincide below $\phi \approx 
0.63$, the $n_r=18$ data 
at large pressures have not reached thermal equilibrium.}
\end{figure}

\begin{figure} 
\resizebox{0.48\textwidth}{!}{\includegraphics{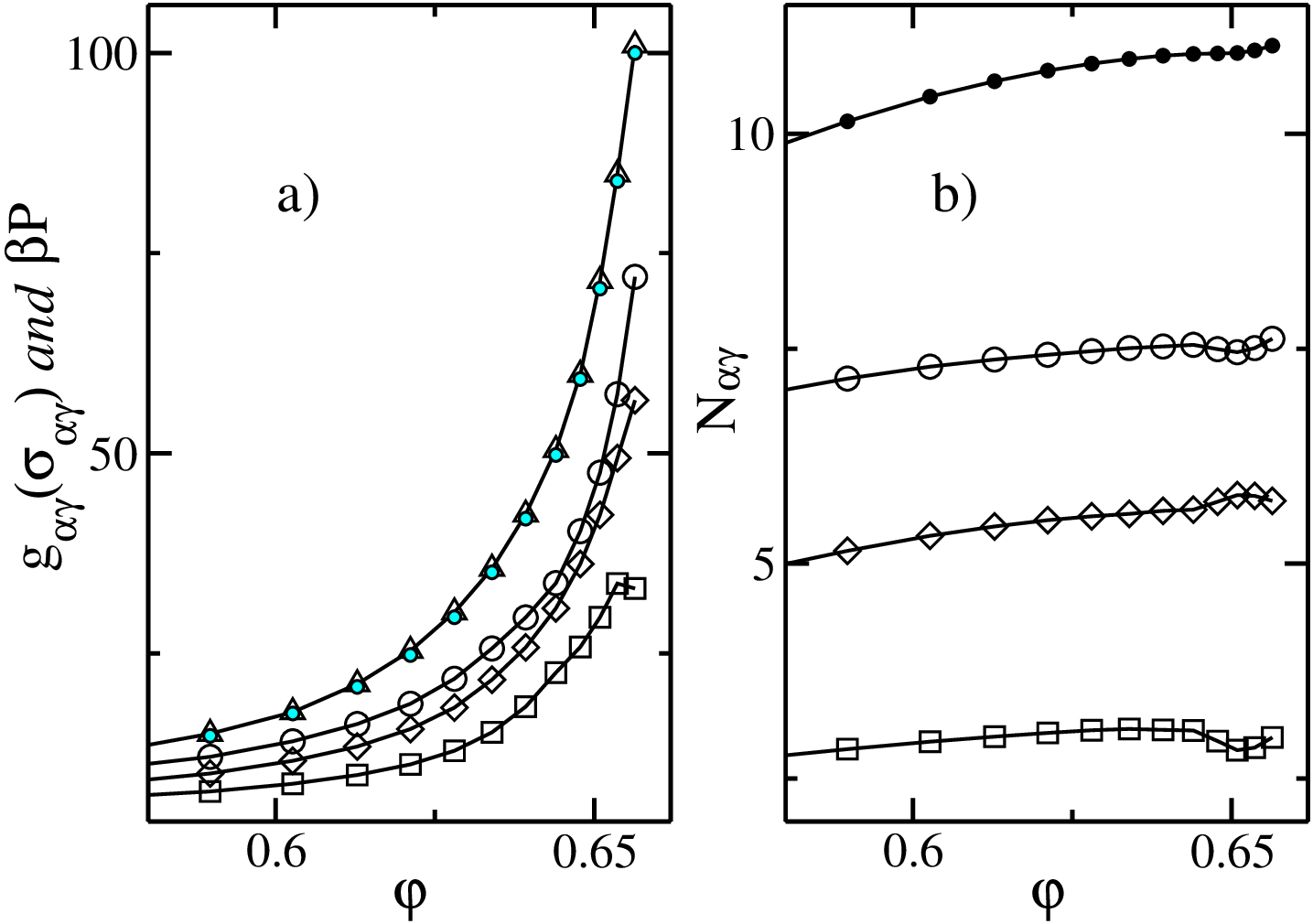}}
\caption{\label{Press-bond-long} Same as Fig.~\ref{Press-bond}
for the run with $N=60$ and $n_r=18$. The data scatter at large volume
fraction, $\phi > 0.63$ indicates nonergodic effects.}
\end{figure}

In the previous section we found the unexpected result that, 
for modest system sizes, equilibrium data could be produced 
even beyond the fitted location of the VFT singularity.
In this section we ask whether it is possible to go to even larger
volume fractions and cross $\phi_0 =0.635$, the putative location of the 
thermodynamic glass transition estimated in Ref.~\cite{Berthier09}.   

To start answering this question we run a simulation with $N=60$ 
and a larger number of replicas, $n_r=18$ setting the maximum
pressure to $\beta P = 100$ and a geometric 
factor of 0.84. As before, we discard the first 
$2 \times 10^{13}$ trials, and use $ 2 \times 10^{13}$ trial moves to 
perform measurements. 

The results for $Z(\phi)$ and $\chi(\phi)$ are shown in
Fig.~\ref{Z-chi-long}, while the pressure, contact values of the radial
distribution functions, and number of neighbors are shown in
Fig.~\ref{Press-bond-long}.
The data shown in Fig.~\ref{Z-chi-long} are consistent with those 
found previously with $n_r=14$. 
There is not only a good agreement with the data obtained for $n_r=14$
and $N=100$, but also with the free volume extrapolation 
towards larger $\varphi$.
Additionally, for $\phi < 0.63$, the measured pressure
shown in Fig.~\ref{Press-bond-long}-a matches the imposed 
pressure and all structural quantities display a smooth evolution 
with $\phi$, see Fig.~\ref{Press-bond-long}-b. 
Unfortunately, this smooth behavior is lost for $\varphi \gtrsim 0.63$, 
see Fig.~\ref{Press-bond-long}, suggesting that 
inadequate sampling is performed. 

\begin{figure} 
\resizebox{0.45\textwidth}{!}{\includegraphics{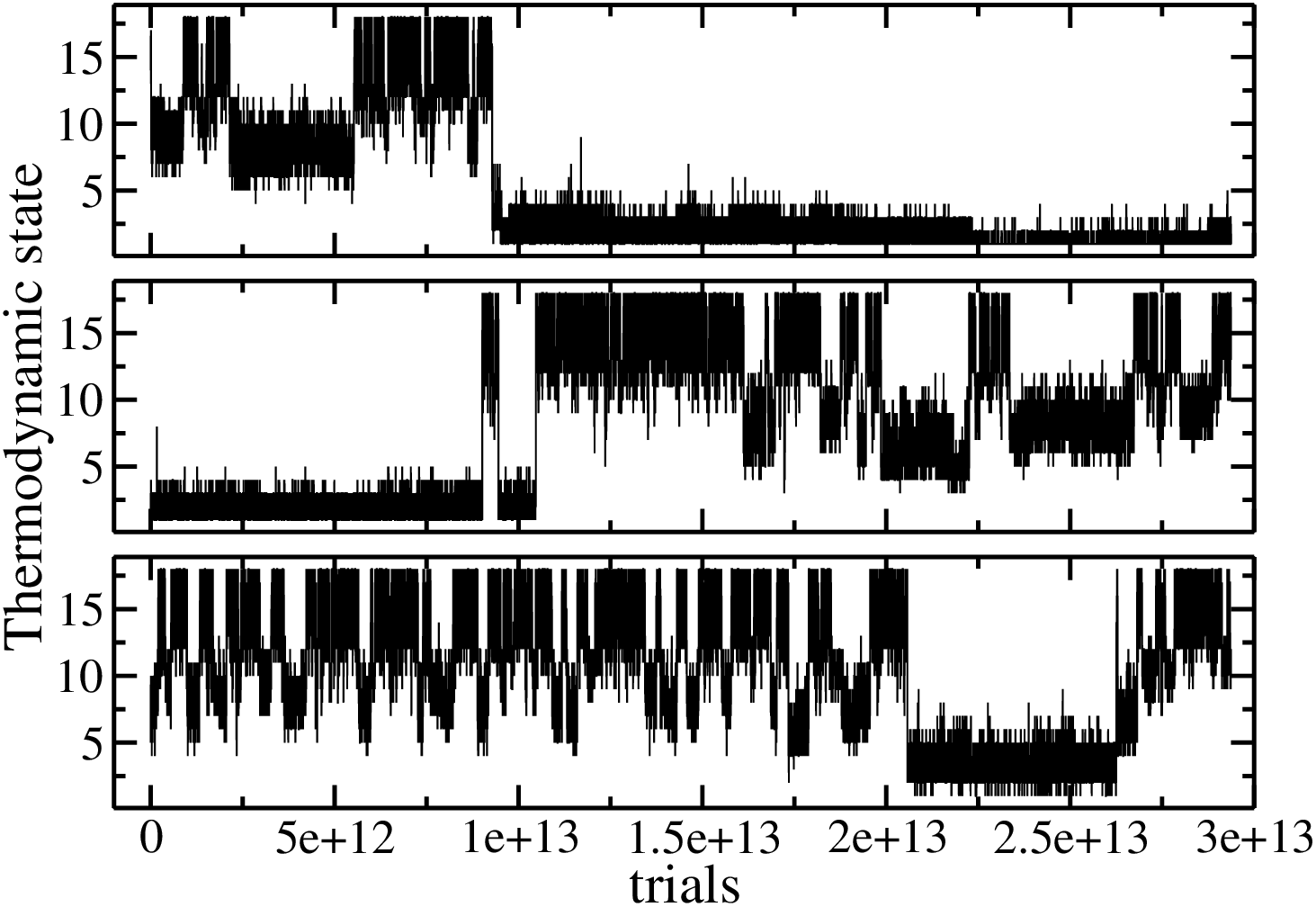}}
 \caption{\label{Swap-long} Same as Fig.~\ref{Swap} for the run 
with $N=60$ and $n_r=18$. Thermodynamic state `1' corresponds to the highest
pressure and `18' to the lowest.
Contrary to Fig.~\ref{Swap} here the replica do not appropriately 
visit all thermodynamic states in an ergodic manner.}
\end{figure}

This conclusion is further supported by the data shown 
in Fig.~\ref{Swap-long} which shows the path in the pressure space of 
three chosen replicas. 
Despite an acceptance rate for replica exchanges being close to $10
\%$, the replicas clearly do not sample all thermodynamic states
with equal probability. In particular, it is clear that the 
averages at large pressure are performed over a very limited number of 
independent configurations, suggesting that an ergodic sampling of the 
phase space is not achieved. 
Note that the third replica, before getting arrested at large pressure
near the end of the run, smoothly travels among the 
highest 14 pressures, which probably explains why the two data
sets with $n_r=14$ and $n_r=18$ produce consistent results 
below $\phi=0.63$, despite the fact that the latter run is clearly not 
producing fully thermalized data. 

Therefore, we conclude that much longer simulations would be needed 
to reach thermal equilibrium above $\phi \approx 0.63$, presumably with 
a larger number of replicas to allow a more extensive sampling of the
configuration space. Unfortunately, this implies
that despite our numerical effort we cannot discuss 
the possibility raised in Ref.~\cite{Berthier09} that a thermodynamic
glass transition takes place near $\phi_0 = 0.635$ in the present
binary hard sphere mixture.

\section{Discussion}
\label{conclusion}

In this work, we have demonstrated that
the replica exchange Monte-Carlo method recently adapted to 
improve the sampling of hard sphere systems is a useful new tool to 
investigate the thermodynamic behaviour of 
the disordered fluid state in a binary mixture of hard spheres
up to very large volume fractions. 
We found that reproducible, thermalized results 
could be obtained up to $\phi \approx 0.63$
at least for moderate system sizes, $N \leq 100$.
This volume fraction is beyond two important `critical'
packing fractions defined dynamically, namely
the mode-coupling transition $\phi_{\rm mct} =0.592$ and 
the divergence extrapolated using a Vogel-Fulcher-Tamman expression,
$\varphi_{\rm vft}=0.615$. Following the equation of state
for the pressure $Z(\phi)$ and the isothermal compressibility
$\chi(\phi)$ we found no signature of a thermodynamic 
glass transition up to $\phi = 0.63$ for the system sizes we use.   
Additionally, we have pushed the lower bound for the location of the
divergence of the pressure of the fluid branch up to 
$\phi = 0.669$, much above the previous determination 
$\phi=0.662$. 

For computational reasons our study was limited both 
in the range of system sizes and of volume fractions for which 
thermal equilibrium could be reached. 
Thus, our results leave open the existence of (at least) three different 
scenarios for the behaviour of the fluid of hard spheres at large volume 
fractions in the thermodynamic limit.   

In a first scenario, we assume that finite size effects
are small and that our data at large pressure above $\phi=0.63$
are nevertheless indicative that no change of behaviour 
is to be expected for $Z$ and $\chi$ even at larger volume fractions,
such that Eq.~(\ref{freefit}) will continue to hold up 
to some $\phi_c$. In this view, $\phi_c$ would represent the 
end point of the fluid branch where the equilibrium pressure 
of the fluid would diverge, while the region $\phi=0.56-0.61$
represents a crossover from the BMCSL equation of state 
to a free volume-like divergence. To prove or disprove this scenario 
is nearly impossible, as one should establish that 
no thermodynamic singularity occurs up to $\phi_c$ in the
thermodynamic limit. It is also natural to expect, 
in this perspective, that the equilibrium 
relaxation time of the fluid would also diverge at $\phi_c$.
This was termed the ``jamming'' scenario
in Ref.~\cite{Berthier09} because it is the diverging pressure
that controls the divergence of the viscosity. 
Note that our results imply that this divergence 
will in any case occur above $\phi=0.669$, 
which is much above the location of the jamming transition (`point J')
at $\phi_J = 0.648$ obtained using purely athermal packing 
preparation protocols~\cite{ohern}. 
Thus, even in the absence of a thermodynamic
glass transition, point J does not control the glass transition 
of hard spheres. 

A second scenario could be that finite size effects 
are severe, that our checks with different 
system sizes are insufficient, and that $N=100$ is still very far away 
from the thermodynamic limit even in the crossover regime $\phi=0.58-0.62$. 
In that case, the crossover region we have described could potentially
become sharper, in the thermodynamic limit, yielding a discontinuity
of the pressure and a jump of the compressibility. This scenario is 
potentially simpler to study numerically based on our work, as 
one should attempt to increase further the range of system  
sizes studied while maintaining thermal equilibrium in the 
crossover regime, an objective that does appear
numerically realistic. 

In a third scenario, our conclusions would continue to hold
in the thermodynamic limit up to $\phi=0.63$, confirming 
in particular that nothing special happens near 
$\phi_{\rm vft} = 0.615$. However, a thermodynamic 
transition could still take place
at larger density, as suggested for instance in 
Refs.~\cite{gio,gio2,Berthier09,Berthier09b}
where a dynamic singularity was located near $\phi_0 = 0.635$. 
Although we found no thermodynamic signature of $\phi_0$ in 
Sec.~\ref{nonequilibrium} we also noticed that our data at these large
volume fraction were not thermalized leaving open the possibility that 
a pressure discontinuity exists at equilibrium. Exploring this third 
scenario would be quite demanding, as one would need to cross $\phi_0$ at
thermal equilibrium for larger systems.
 
To conclude, it should come as no surprise that 
providing solid conclusions regarding the existence of a thermodynamic 
liquid-glass transition in the thermodynamic 
limit is a difficult numerical task. However, we have provided evidence
that replica exchange Monte-Carlo simulations can be used 
to study this issue in hard sphere systems, and we have 
suggested that drawing some firm conclusions is perhaps not 
completely out of reach. In particular, it would be interesting 
to study larger system sizes, together with a larger number of replicas
to maintain the acceptance rates for replica exchanges at an acceptable 
level. This implies using larger computational resources.

\bibliographystyle{prsty}

\bibliography{HS-2S}

\end{document}